\documentclass{article}

\usepackage{arxiv}

\usepackage[T1]{fontenc}    
\usepackage{hyperref}       
\usepackage{url}            
\usepackage{booktabs}       
\usepackage{amsfonts}       
\usepackage{nicefrac}       
\usepackage{microtype}      
\usepackage{biblatex} 
\usepackage{pdfpages}
\usepackage[section]{placeins}
\usepackage[bottom]{footmisc}
\usepackage{tikz}

\bibliography{bibtex}

\title{Measuring the Effectiveness of Digital Hygiene using Historical DNS Data}

\author{
  Oliver ~Farnan\\
  University of Oxford\\
  \texttt{oliver.farnan@cs.ox.ac.uk} \\
   \And
 Gregory ~Walton \\
 Oxford Internet Institute\\
 \texttt{gregory.walton@oii.ox.ac.uk} \\  
   \And
 Joss ~Wright \\
 Oxford Internet Institute\\
 \texttt{joss.wright@oii.ox.ac.uk}
}

\begin{document}
\maketitle

\begin{abstract}
This paper describes an ongoing experiment evaluating the efficacy of a digital safety intervention in six high-risk, low capacity Civil Society Organisations (CSOs) in Central Asia. The evaluation takes the form of statistical analysis of DNS traffic in each organisation, obtained via security tools installed by researchers. 

The hypothesis is that the digital safety intervention strengthens the overall digital security posture of the CSOs, as measured by number of malware attacks intercepted by a cloud-based DNS firewall installed on the CSOs networks. 

The research collects DNS traffic from CSOs that are participating in the digital safety intervention, and compares a treatment group consisting of four CSOs against DNS traffic from a second group of two CSOs in which the intervention has not yet taken place. 

This project is ongoing, with data collection underway at a number of Central Asian CSOs. In this paper we outline the experimental design of the project, and look at the early data coming out of the DNS firewall. This is done to support the ultimate question of whether DNS data such as this can be used to accurately assess the efficacy of digital hygiene efforts.

\end{abstract}

\keywords{DNS \and Malware \and Digital Hygiene \and Threat Intelligence \and Network Monitoring \and Network Forensics}

\section{Introduction}
The Cyber Security Audit and Remediation (‘CyberSAR’) project is a two-year effort that offers digital safety research and support for high-risk/low capacity civil society organizations across the Eurasia region. CyberSAR aims to measurably reduce the cyber vulnerabilities of selected high-risk organizations through a combination of security audits, real-time remediation of threats and risks to communication devices, extended user training, policy support, and the deployment of a technical threat mitigation system: a cloud-based DNS firewall.

To evaluate the effectiveness of the CyberSAR intervention, this research describes a time series analysis of the Domain Name Service (DNS) traffic originating from each field location to determine the extent to which security practices have reduced traffic to known malware domains.

Like many digital safety projects in the civil society sector, the CyberSAR intervention operates with constrained resources, limiting costly or complex technical controls. DNS is an essential and lightweight protocol, and it is hoped that efficient analysis of this traffic will give some of the insights gleaned from more comprehensive network monitoring. While DNS is used for everyday benign internet use such as web browsing or certificate validation it is also used by bad actors. A recent survey by Cisco showed that 91.3\% of malware uses DNS in attacks, either to receive instructions from Command and Control (C2) servers or for data exfiltration \cite{Lystrup2016-jd}. 

The same survey indicated that 67\% of organisations do not monitor recursive DNS to detect anomalies associated with known or unknown cyber threat behaviours. Increasingly however, network defenders are monitoring DNS traffic and applying static blacklists of malicious domains to block communication with malware.

In this study we aim to explore whether DNS logs are sufficient to determine the efficacy of digital hygiene treatments. Within our data set there are several organisations that have undergone such treatments, as well as organisations that have not. We hope to do this by studying whether these organisations show significant behavioural differences with their interaction with malicious domains.

The first part of our question asks whether is it possible to assess the efficacy of the digital hygiene remediations using just DNS data. DNS data is useful as it is commonly available and cheap to record and store. While DNS data is single dimensional it is fairly comprehensive, giving a record for all connections established using the domain name system. In many cases historical DNS records are likely to exist even for organisations that have no formal security logging and monitoring in place.

The second part of our question asks that if it is possible to assess the efficacy of a digital hygiene campaign, has this particular digital hygiene campaign had any observable impact? This is more difficult to show, as it depends on what extent we're able to answer the first question, and it is notoriously difficult to assess the value of cyber security controls.  

\FloatBarrier

\subsection{CyberSAR Process}

To understand what we are testing it is important to understand the digital hygiene process we are looking at. This process has 8 steps, and is referred to as the CyberSAR process. They involve a series of steps provided by in country audit teams. The priority is to help beneficiaries protect their data, identities and operations online, and to confidently use communications devices. The project also seeks to generate and disseminate evidence-based research on the Eurasian cyber risk/threat environment for civil society in the region and disseminates cyber security alerts, bulletins and technical advice. The cloud-based DNS firewall is installed in advance of this process, once organisational consent has been obtained. The duration of the experiment is 2 months, although the CSOs may opt-in to firewall coverage beyond this.

(a) pre-audit survey: once the CSO has agreed to participate in the audit process they complete a pre-audit survey. This information is to familiarise the audit team with the CSOs digital security capacities and priorities.

(b) attack surface analysis: this social vulnerability assessment maps the CSOs online presence, including employees use of social media, to determine if private or sensitive information is being exposed which, in the hands of an adversary, could be weaponised - for example, to craft a lure for a targeted malware attack. 

(c) simulated phishing: a simulated spear phishing link is sent to employees to quantify the CSO’s vulnerability to phishing attempts. This is followed up with hands-on training.

(d) on-site assessment and real-time remediation: the audit team conducts an on-site assessment of organisational digital security practices and procedures. Key vulnerabilities are addressed through hands-on remediation.

(e) post-audit discussion: the site visit concludes with a verbal briefing of the main findings, highlighting any critical digital security risks and sketching-out follow-up action. 

(f) final report and recommendations: a concise report describing the current state of the CSO’s cyber security. The report identifies the most important digital safety deficiencies and provides practical recommendations to resolve them, as well as a follow-up support and training plan.

(g) post-audit remediation and training: in the period after the audit, the field team work with the CSOs to resolve security issues highlighted in the report and to train staff to implement good digital security practices and develop simple organizational policies. 

(h) six-month check-in: approximately six months after the initial audit a follow-up visit to the CSO is scheduled to determine whether key digital security policies and practices are still in place and that good digital hygiene is being practised by employees.

Having summarised the 8 steps of the audit and review process, the efficacy of which we will be evaluating, there are a number of features in the technical system design to ensure the confidentiality, integrity, and availability of the data throughout the experiment \cite{Andress2014-yy}.

Firstly, the primary protocol on which we are collecting data is DNS. DNS is an unencrypted protocol, with queries sent in plain text to the ISP's servers. Thes can therefore be read, altered, or blocked by either the ISP, an intermediary on the line, or anyone with authority over the in-country network. 

The DNS firewall used in this research encrypts DNS traffic as it traverses the network. In contrast similar services that use DNS resolvers to enhance digital protection do not routinely offer this protection. Third-party DNS resolution services such as Google, Quad9, Cloudflare, and others send the organisations' DNS data via unencrypted channels, although support for encrypted DNS (via DNS over HTTPS or DNS over TLS) is increasing. Participants in this research are provided with the OpenDNS DNSCrypt utility to connect to the DNS resolver, and therefore benefit from the encryption of queries. This assures the confidentiality and integrity of resolution data, and guards against DNS spoofing attacks \cite{Maksutov2017-pe}. 

Secondly, regarding the confidentiality and integrity of our DNS resolvers, each node in the DNS resolver infrastructure is hardened against attack by running only the services required to resolve and log queries. All unnecessary services are turned off and ports closed. Each query is sent by beneficiaries to a load-balancing node running DNSDist\cite{Sousa2017-ha}. 

This DNS load-balancing node is hosted on Amazon Web Service's infrastructure in Frankfurt. Each load-balancing node only accepts incoming queries that use the correct cryptographic key, or that originates from an expected IP range. Queries are then forwarded from these nodes to DNS resolvers, which use IP whitelisting to ensure that only a specific set of IP addresses may query it. All other incoming communications are blocked. 

Each resolver stores logs of DNS queries for a very short period as they are sent to be processed and stored on AWS' S3. The treatment group's actual IP address is not logged during this process and they can only be identified through their assignment to a load-balancing node which is used exclusively by them. 

Thirdly, with regards to the logging infrastructure, each query is logged to Amazon S3, which can only be accessed using keys generated and stored securely by authorized users. Finally, a collaborative analytic dashboard is built on the Elastic stack \cite{Gupta2017-vg} a distributed, JSON-based search and analytics engine, which offers authentication, authorization, encryption of data, as well as audit logging. The visualisation of the data in Elasticsearch is performed using Kibana \cite{Gupta2015-ci}- as extensible user interface for configuring and managing the Elastic stack. 

\FloatBarrier
\section{Experiment and Data}

Our analysis of the digital hygiene efforts was based on two data sets: a DNS results queries dataset, and a threat intelligence based dataset listing suspected malicious domains. Both of these were taken from the DNS firewall used by the CSOs.

\subsection{DNS Queries}

\begin{figure}
\centering
\includegraphics[width=\textwidth,height=\textheight,keepaspectratio]{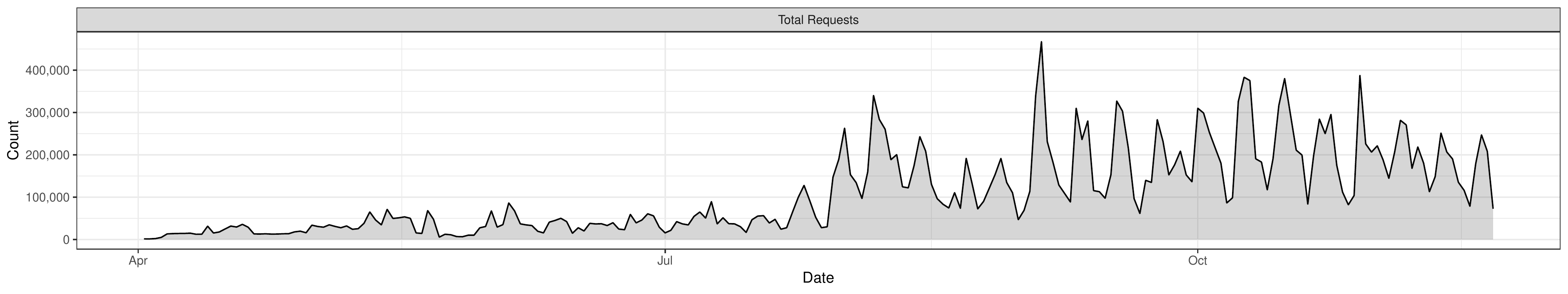}
\caption{Total observed requests over time}
\label{fig:data}
\end{figure}

This initial research report is based on the analysis of DNS data from six different CSOs taken between April and November 2018. This data contains queries for 60,000 individual domain names totalling 8,000,000 DNS queries. At its peak there are over 450,000 queries made in a single day. The data is spread over 8 months, and includes the majority of DNS queries these organisations made\footnote{There is some occasional use of USB internet dongles that set an alternative DNS address.}. A time series representation of this data can be seen in Figure ~\ref{fig:data}.

Four of these six organisations have received the digital hygiene efforts described above, and these groups are collectively called the Treatment group. The other two organisations have not, and are collectively described as the Control Group. Where we break out the individual organisations for analysis we describe them by the colours of their plots: Red and Yellow are Control organisations, while Green, Turquoise, Blue, and Pink sit within the Treatment group.

Within this report we have only analysed the first eight months of data. Data has been collected since November 2018, and is being collected going forward. This further includes organisations other than the six studied here, and crucially for future analysis, contains organisations switching from the Control to Treatment groups.

\FloatBarrier
\subsection{Threat Intelligence}

The DNS service includes the functionality of a DNS firewall. This allows the centralised blacklisting or alerting when queries are send for suspicious or known malicious domains.

This data was was made available to us for this analysis, and consists of over 300,000 domains and subdomains. The vast majority of entries on this list came from open source threat intelligence feeds, although the platform also allows analysts to manually flag domains. Most of these entries also have corresponding tags such as `malware' or `phishing', to give some idea to the purpose of the inclusion of the domain.

Although this list contains 300,000 domains the majority of the these cannot be considered malicious without further individual examination. Most of the domains on the list are benign, including at least several domains that appear to be incorrectly flagged as malicious. This could have occurred from errors in threat intelligence feeds or in response to an attack coming from a normally benign domain.

\begin{figure}
\centering
\includegraphics[width=\textwidth,height=\textheight,keepaspectratio]{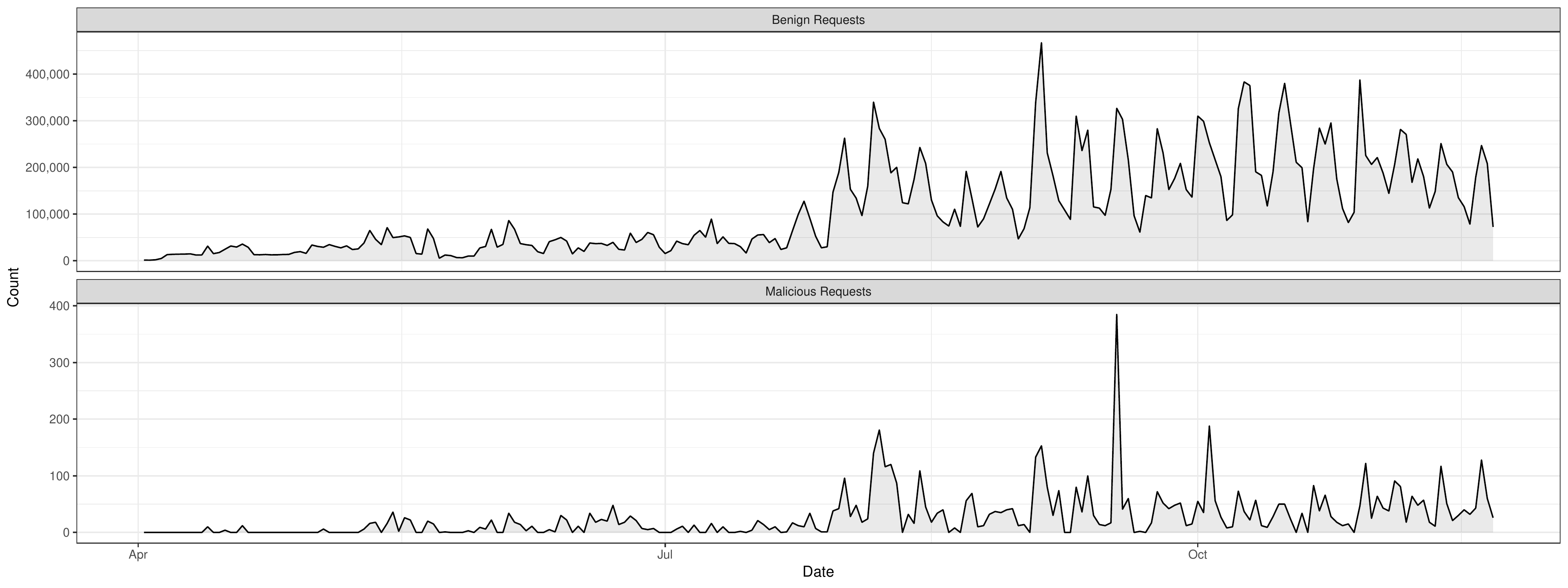}
\caption{Absolute number of benign and malicious requests}
\label{fig:malicious}
\end{figure}

\begin{figure}
\centering
\includegraphics[width=\textwidth,height=\textheight,keepaspectratio]{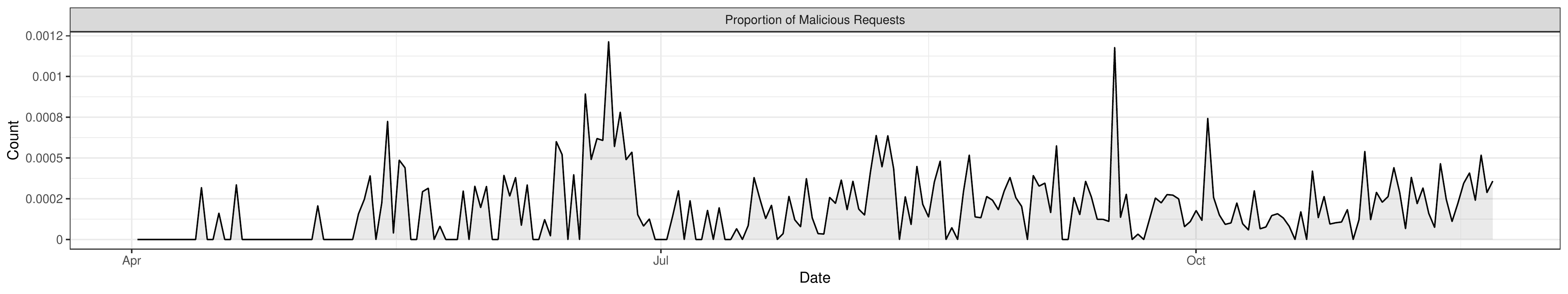}
\caption{Proportion of requests which are malicious}
\label{fig:malicious_proportion}
\end{figure}

Figure ~\ref{fig:malicious} shows the total number of queries made within the period that are classified as malicious according to the threat intelligence list. These are domains which have been set to either be Blacklisted or Convicted by the DNS firewall, and are tagged with tags such as 'botnet`, 'malware`, or 'virus`. We can see that the number of both malicious and benign requests grow over time. Figure ~\ref{fig:malicious_proportion} shows these queries as a proportion of the total queries for any given day. 

The number of malicious requests fluctuates, and in particular there are several sharp spikes, the biggest of which occurs on 2018-09-17. Unfortunately this is spike appears to be caused by one user seeking torrents and pornography: there are repeated requests for sites such as chaturbate.org, cams.com, utorrent.com, and mininova.org. These requests span over five hours, indicating it took the user awhile to find what they were looking for.

While this noise in the data is frustrating it is demonstrative of the 
difficulty in automating the process of attack discovery from DNS data. According to these data sets 2018-09-17 saw double the hostile domain requests of any other single day, yet upon investigation this appears to be a red herring. While quantitative time series analysis is insightful the numbers can be misleading. It also raises the question of the usefulness of threat intelligence feeds without manual investigation.

\FloatBarrier
\section{Initial Findings}

\subsection{Most Frequent Requests}
\begin{figure}
\centering
\includegraphics[width=\textwidth,height=\textheight,keepaspectratio]{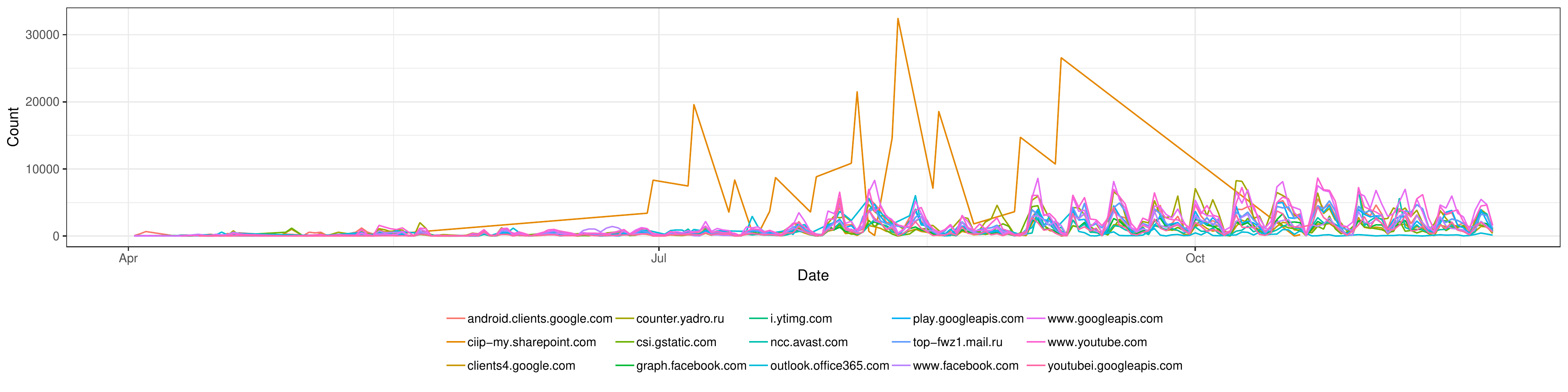}
\caption{List of most commonly searched for FQDNs}
\label{fig:combined_domains}
\end{figure}

The most commonly referenced fully qualified domain names (FQDN) can be seen in Figure ~\ref{fig:combined_domains}. While most of these domains follow a gradual increase and weekly pattern there is one domain that stands out: ciip-my.sharepoint.com. Requests for this domain are not consistent, but when they do occur they often occur in large numbers.

The ciip-my.sharepoint.com subdomain is only actually accessed on 30 days throughout the observation period. However, on days when it is accessed it is frequently accessed thousands or tens of thousands of times. Over 99\% of the requests for this subdomain come from a single organisation: presumably the organisation to which the sharepoint belongs. It is likely that these sporadic high-volume queries are caused by poorly programmed software resulting in repeated requests to this single subdomain.

\begin{figure}
\centering
\includegraphics[width=\textwidth,height=\textheight,keepaspectratio]{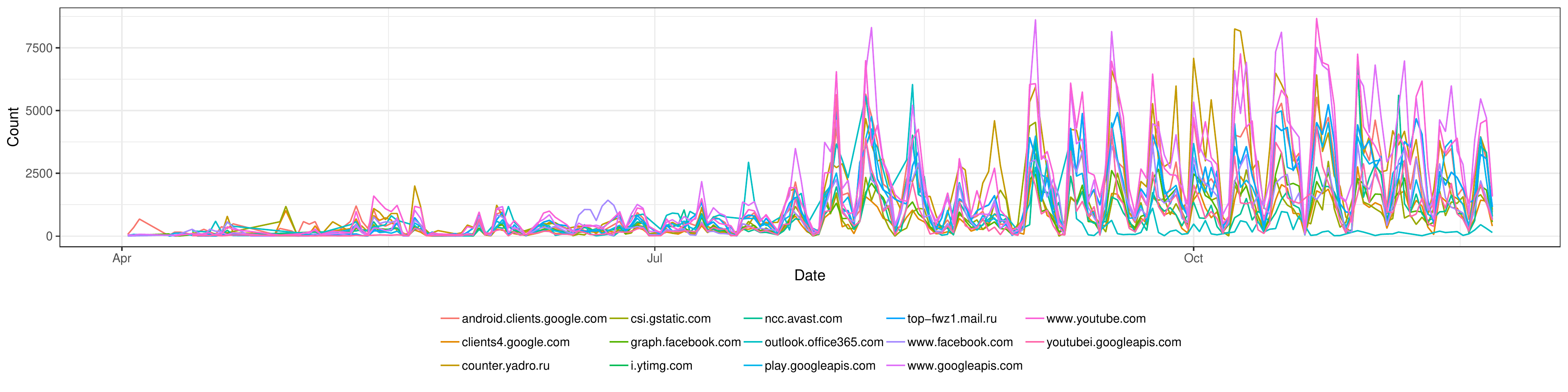}
\caption{List of most commonly searched for FQDNs with anomalous entry removed}
\label{fig:combined_domains_wo_ciip}
\end{figure}

Figure ~\ref{fig:combined_domains_wo_ciip} has this domain removed, and shows a more regularly weekly pattern for most of the top domains. There are still some fluctuations however, for example outlook.office365.com is in regular use for the first half of the data set, but then its use drops to a low level mid September. The majority of outlook.office365.com requests were from a single organisation, and it is likely that the organisation switched to a different platform during the observation period.

\begin{figure}
\centering
\includegraphics[width=\textwidth,height=\textheight,keepaspectratio]{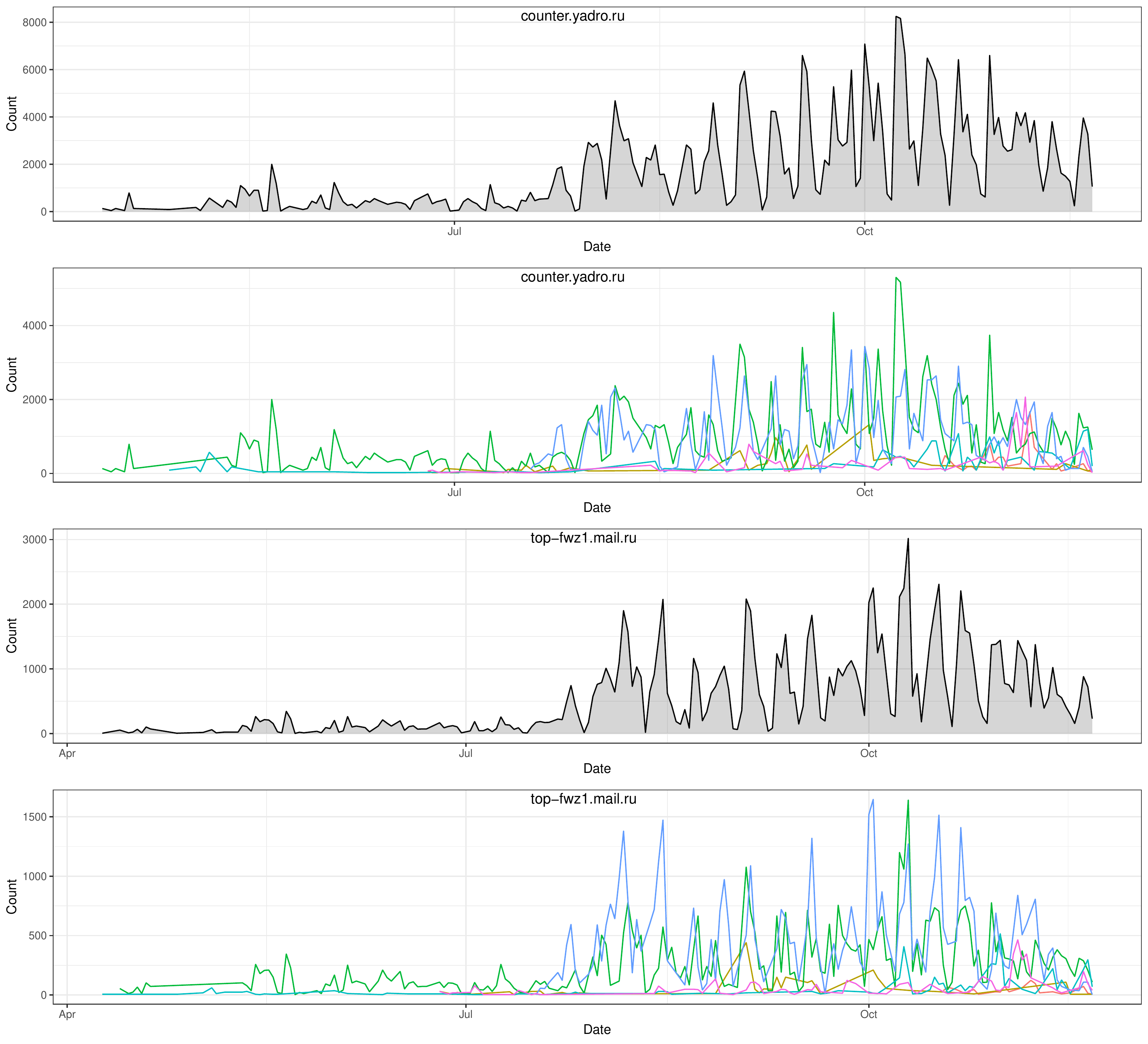}
\caption{Requests for counter.yadro.ru broken and top-fwz1.mail.ru broken down by organisation}
\label{fig:yadro-top-fwz1-combined}
\end{figure}

Just amongst these top 15 domains there are two malicious associated domains making an appearance. counter.yadro.ru and top-fwz1.mail.ru are low quality advertising domain known for accepting ad-hijacking requests and malvertising. Figure ~\ref{fig:yadro-top-fwz1-combined} shows the frequency with which these domains are accessed, in total and broken down by organisation. These domains are frequently visited throughout the entire observed period commensurately with the overall increase in traffic. They also shows signs of weekly usage patterns.  counter.yadro.ru is very popular, and for a week in August and several weeks in October this domain is actually the most frequently accessed domain of any in the data set.

Two organisations in particular appear to be making the majority of requests to these adware domains. While one of these organisations (Green) makes up most of the requests to counter.yadro.ru and the other (Blue) makes up most of the top-fwz1.mail.ru, there is a significant amount of overlap between the two. A third organisation (Pink) appears to have a spike towards the end of the dataset, making requests to both domains, but this appears to be mostly cleared up after a week.

There are two likely causes for this pattern, both of which are associated with poor digital hygiene. The first is that these queries are coming from hosts infected with malware or potentially unwanted programs. These domains are associated with ad hijacking, where the users' web ads are redirected to ads that profit the hijacker, and there is little reason for users to browse to these domains directly. This malware is unlikely to be targeted, and we cannot assume that these hosts are infected with more insidious or powerful malware, such as botnets or backdoors.

The other option is that users are browsing to risky domains that serve ads from these networks. Risky sites such as pornography and file sharing often do not comply with the stricter criteria of more respectable online advertisers, and resort to low quality advertisers.

The requests follow the pattern of a five day working week, indicating that the affected machines are user workstations rather than high-uptime servers. While these domains are not a direct indication of attack or compromise, they do give an indication of the digital hygiene and overall security posture of the organisations.

Both of these domains are on the malicious domains list, but are tagged as adware and spyware, and not malware or malicious. These domains introduce the concept of grey domains, which we explore further shortly.

\FloatBarrier
\subsection{Break down by different organisations}

\begin{figure}
\centering
\includegraphics[width=\textwidth,height=\textheight,keepaspectratio]{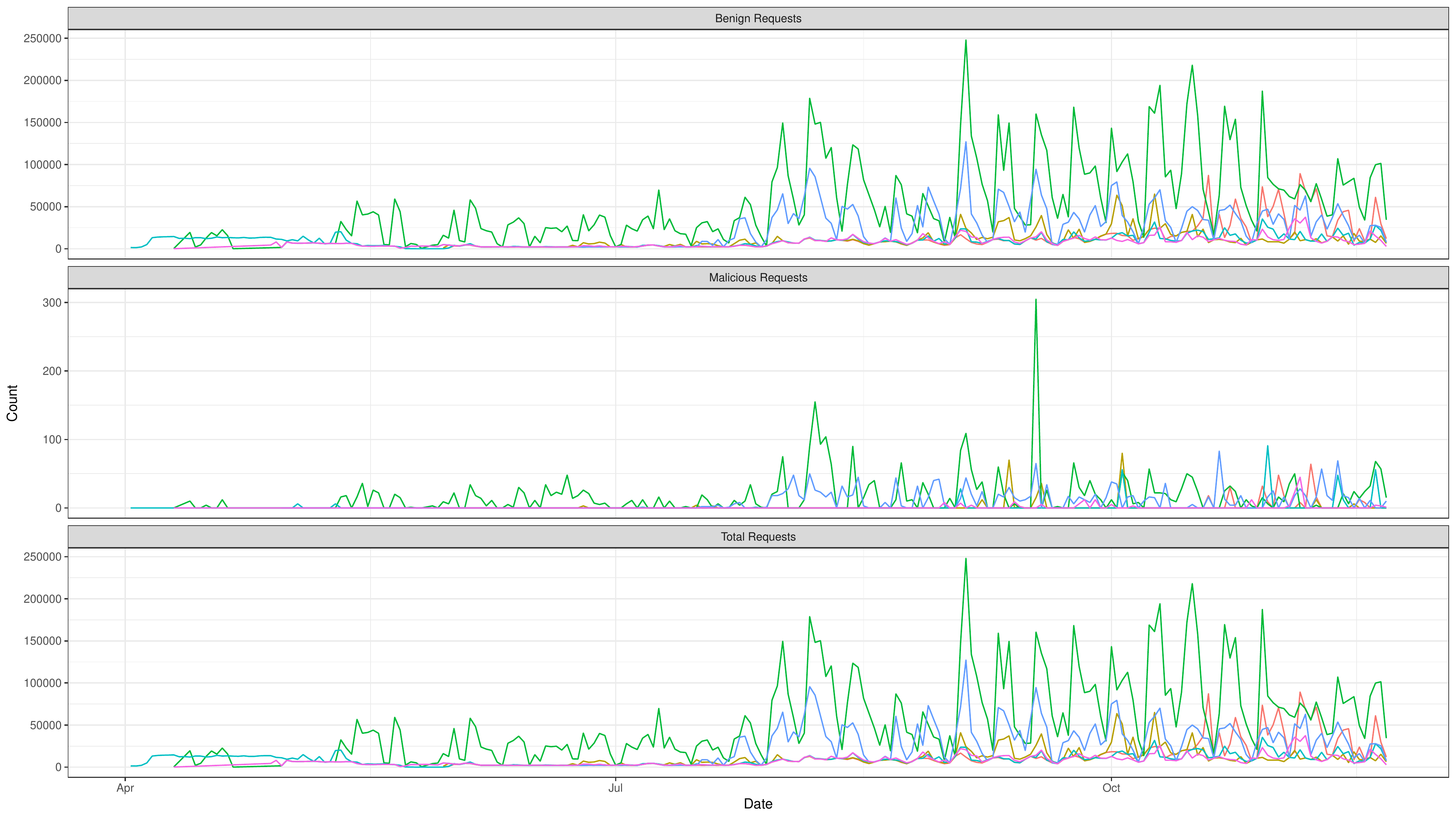}
\caption{Benign, Malicious and Total requests, broken down by organisation}
\label{fig:different_orgs_summary}
\end{figure}

There are six different organisations that we have full data for over the period. Figure ~\ref{fig:different_orgs_summary} shows the benign, malicious and total requests for these separate organisations.

When compared to only looking at the overall spikes in the previous section, we now see that most spikes are caused by individual organisations. In particular Green and Blue seem to experience a high number of potentially malicious requests. We covered the spike on 2018-09-17, but it is now apparent that Green has other spikes during 2018-08-07 to 2018-08-09, and 2018-09-03 to 2018-09-04.

The Green spikes between 2018-08-07 to 2018-08-09 consist largely of requests to www.odnoklassniki.ru (a Russian social network) and dbk589trlnxim.cloudfront.net (a site advertising Android gaming emulation software). The spike on 2018-09-03 to 2018-09-04 appears similar, again with lots of requests to www.odnoklassniki.ru, mixed in with an assortment of torrent sites.

When looking at the proportion however these spikes are not disproportionate. Green is busy those days in general, and the spikes in these `malicious' domains is commensurate with that use. This kind of traffic is common for Green over the full time period - at least during the working week.

Looking at these domains it is clear that what we're seeing with Green isn't evidence of a targeted attack, but they have at least one user who often visits domains that have been associated with malicious use in the past. This user is connected to the network during the week, and likes to visit www.odnoklassniki.ru, torrent sites, and pornography while connected to the work network.

There are two interesting discoveries from this:

Firstly, it is clear that we must be sceptical of the `malicious' domains within the threat intelligence feed. Pornographic and torrents sites are often linked to malicious behaviour, but it is typically through malvertisement and normally fleeting. Sites such as utorrent.com and chaturbate.org are major websites and highly frequented domains, that are not serving malicious content to the majority of their users. Should these domains be flagged as malicious? That is questionable. They contain a high trust score on trust sites such as virustotal.com, and without further information it is unclear why they are listed as they are.

Secondly, despite that scepticism, this pattern of behaviour is enlightening about the organisation's digital hygiene practice. An organisation that has users who frequent pornography and torrent sites, through the work network and during the working week, is not operating at with a high degree of digital hygiene. This in itself is telling, and despite being a Treatment organisation, Green in particular seems to have at least one problem user.





\FloatBarrier
\subsection{Grey Domains}

As well as malicious domains, the threat intelligence data set includes information about domains that while not unambiguously malicious, may be unwanted or undesirable. Examples of these domains include web trackers, low quality advertisements, and potentially unwanted applications. While the presence of these domains is not directly linked to malicious use, they are unlikely to be accessed intentionally, and may be a sign of poor digital hygiene practices. We will refer to these domains as grey domains.

One of the problems with using historical DNS queries to assess digital hygiene is that purely malicious domains are often short lived. Regardless of whether targeted or indiscriminate, malicious domains often aim to operate in some level of secrecy. When they are identified they may be blocked or forced to use alternative domain names. As a result malicious domains are constantly changing, and rarely have a long operational life. 

Because of this, threat intelligence lists of malicious domains must too be constantly updated and can never be exhaustive. Threat intelligence feeds are in a constant red queen race to keep up to date with the latest malicious domains. And even if a malicious domain is discovered and blocked by \textit{someone}, this knowledge may not propagate out to other threat intelligence lists, ensuring it is impossible to collate comprehensive lists of historical malicious domains. Fast flux domain changing adds to this problem.

Grey domains do not suffer this penalty. While they are rarely intentionally visited, they are less frequently blocked as they do not cross the boundary into the outright hostile. As a result they less frequently have to switch to alternative domains, and so threat intelligence feeds can include them with more consistency.

\begin{figure}
\centering
\includegraphics[width=\textwidth,height=\textheight,keepaspectratio]{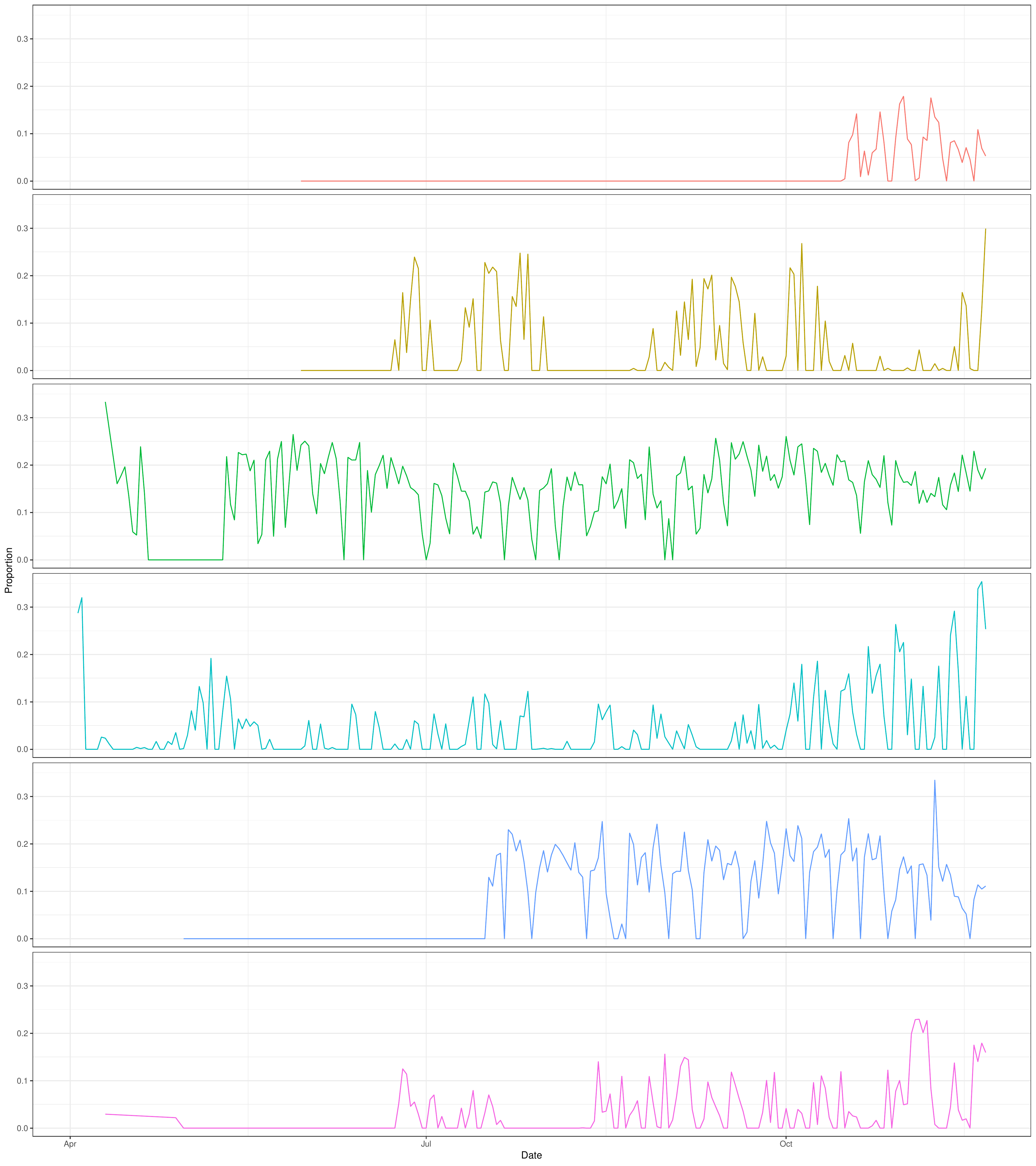}
\caption{Proportion of grey requests broken down by organisation}
\label{fig:grey_orgs_2}
\end{figure}

We split out requests for grey domains and compared them against the proportion of total requests per organisation. This is a noisy graph, so in Figure ~\ref{fig:grey_orgs_2} we display each organisation on its own plot, to allow easy identification and comparison across the six different organisations.

All of the organisations within the experiment show some signs of grey domain activity, although in most cases this is periodic. Several of the organisations show clear weekly patterns for accessing these domains, with breaks during the weekend. This pattern is especially strong for Green, Turquoise, and Blue, and is similar to the pattern we saw with some of the malicious domain requests. What is different is that Figure ~\ref{fig:grey_orgs_2} shows the \textit{proportion} of grey requests increases during the working week, and not just the absolute amount\footnote{If you compare this to the proportion of calls in Figure ~\ref{fig:malicious_proportion} the weekly pattern is less apparent.}. We can see that there is a level of constant DNS network activity that is occurring even during the weekends which does not always include a large number of requests to grey domains, but during the working week these requests increase.

Green and Turquoise are the only organisations to make grey requests over the entire period. The other four organisations start later in the observed period, and also have periods of quiet part way through. There are a number of possible causes for these lulls, and it may be possible that these organisations are not routing all of their traffic through the DNS firewall for the entire period. This will require further investigation in the next stage of analysis.


\FloatBarrier
\subsection{Control vs Treatment groups}

As mentioned in section 2, four of the organisations had received digital hygiene efforts, while two of them had not. While we should be cautious about treating this division as a scientifically fair direct comparison, it does allow us to explore the differences between them.

\begin{figure}
\centering
\includegraphics[width=\textwidth,height=\textheight,keepaspectratio]{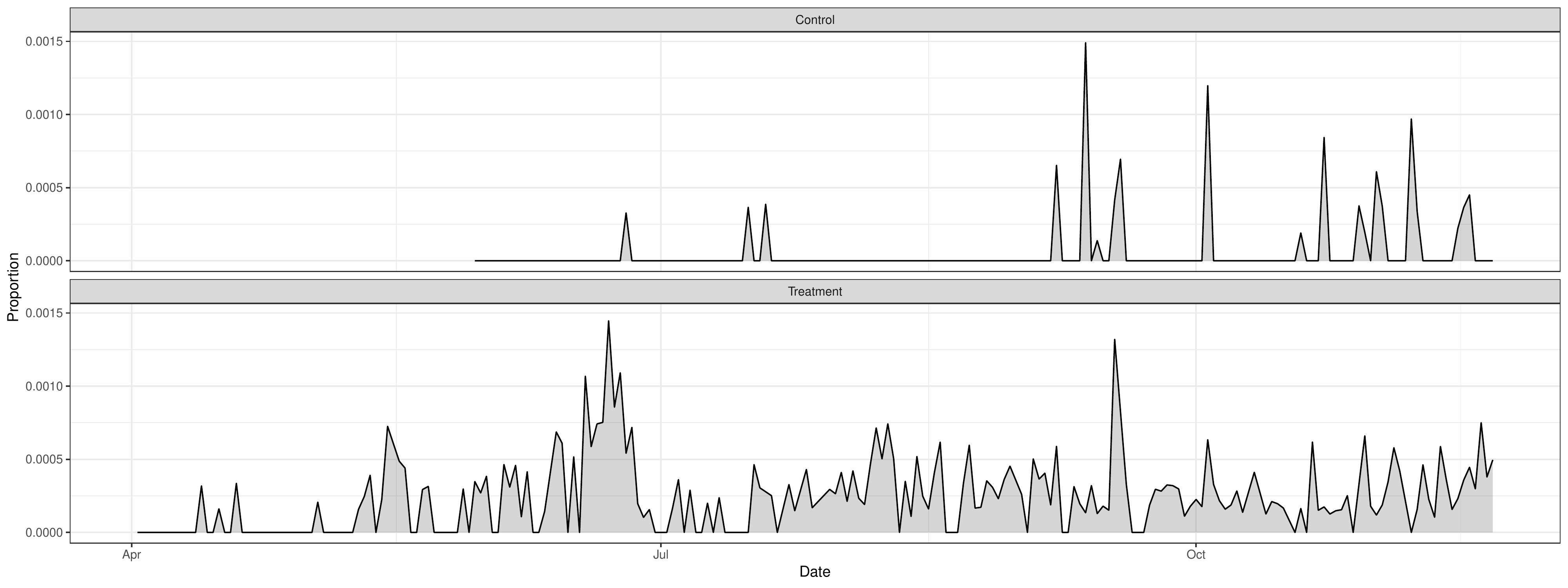}
\caption{Proportion of Malicious requests broken down by Treatment and Control}
\label{fig:cvt-malicious}
\end{figure}

The proportion of malicious requests over the Control and Treatment organisations can be seen in Figure ~\ref{fig:cvt-malicious}. Perhaps surprisingly, the Control group consistently suffers from fewer requests to malicious domains than the Treatment group. The Treatment group maintain a regular pattern of requests to malicious domains over most of the observed period, while the Control group's requests to malicious domains are more sporadic. For most of the observed time period the Control group do not seem to make any requests that have been flagged as malicious from the DNS firewall and threat intelligence feeds.

\begin{figure}
\centering
\includegraphics[width=\textwidth,height=\textheight,keepaspectratio]{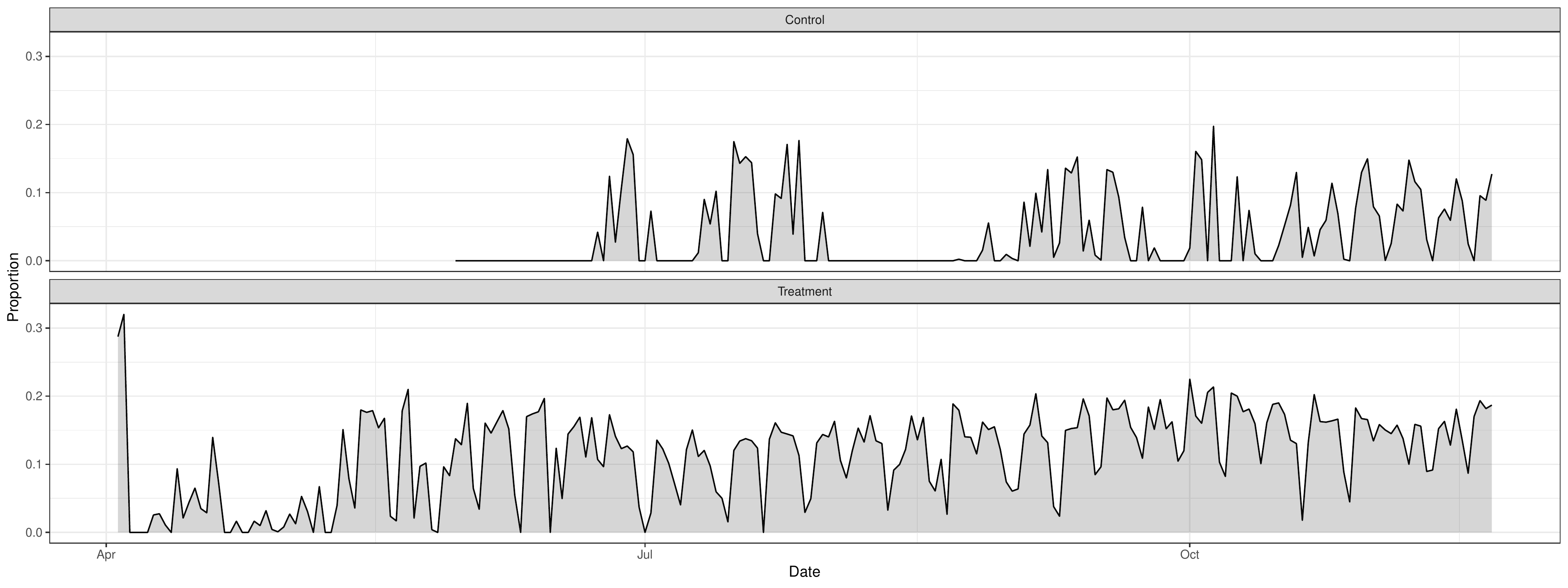}
\caption{Proportion of grey requests broken down by Treatment and Control}
\label{fig:cvt-grey}
\end{figure}

This trend appears to continue with requests to grey domains, which can be seen in Figure ~\ref{fig:cvt-grey}. While these requests are more frequent than malicious requests, the Control group consistently makes less grey requests than the Treatment group. These requests occur over the entire time period for the Treatment group, whereas the Control group suffers from a smaller number of isolated incidents.

As discussed above, the grey requests more closely fit the pattern of a working week than the malicious requests. While malicious requests occur with a greater degree of fluctuation, grey requests appear fairly consistent. This is especially true when looking at the grey requests made by the Treatment group in Figure ~\ref{fig:cvt-grey}. Referring back to section 3.3, we can see the infrequent grey requests made by the control group shown as the Red and Yellow organisations in Figure ~\ref{fig:grey_orgs_2}.

It is counter intuitive that the Treatment group has a higher proportion of malicious and grey requests than the Control group. The digital hygiene efforts (outlined in section 1.1) should have made the organisations less likely to visit domains with these negative tags. But instead they consistently have more requests to these domains, both in terms of absolutes and as a proportion of total requests.

One possible explanation is that that these are smaller organisations with a smaller number of users. Looking back at Figure ~\ref{fig:different_orgs_summary} we can see that both of the Control organisations (Yellow and Red) do not have many total requests compared to the other organisations. The users they have may be more expert users who are less likely to visit malicious domains, or it may be that the larger organisations are disproportionally affected by a few `bad egg' users. When looking at the malicious spikes we saw that often they appeared to be caused by a single user behaving badly, and it is possible that these users are a strong enough influence to skew the whole organisation for the entire time period.


Clearly this needs to be explored in more detail. In the next section we are going to look at our next steps, and what possibilities we have to explore the causes of this contradictory finding.

\FloatBarrier
\section{Discussion}

The purpose of this report is to make an assessment of the possibility of assessing the efficacy of digital hygiene efforts by analysing the DNS data of different organisations. We are basing this analysis on the DNS data of six different CSOs spanning eight months, four of which have received digital hygiene, and two of which have not.

We have used a collation of open source threat intelligence feeds to give information on the domains within the DNS data set. This threat intelligence data set is the same set that is used by the DNS firewall providing monitoring and filtering for these organisations.

At this stage of our investigation we are able to draw four key findings:

\textbf{The data set gives an indication of digital hygiene practices, and consistent patterns can be observed over the whole period, over both malicious and grey domains}. The Green and Blue CSOs suffer particularly badly from this, even when we look at those requests as a proportion of the total. The high number of requests suggests these organisations may have more users than some of the others, and it seems that some of these users are not following good digital hygiene practices. Manual investigation of these organisations confirms that poor user practice is responsible for much of this, and this could be addressed by digital hygiene efforts.

\textbf{\textit{However}, much of the Threat Intelligence flagging is questionable without manual investigation.} Manual investigation reveals that many of the domains that show up as either malicious or grey are probably not acting in an actively malicious way at the time when the domain is accessed. While this kind of manual investigation is acceptable in the context of a a Security Operations Centre, it limits our ability to systematically draw rigorous conclusions. While the data is sufficient to give an indication of the state of digital hygiene of an organisation, it is not possible to make further assessments about specific attacks without more accurate threat intelligence.

\textbf{Is systematic quantitative evaluation useful without manual investigation? Probably.} Despite the questionable nature of some of the flagging within the threat intelligence data set, we believe that the two data sets combined do give valuable insight into the digital hygiene practices of the CSOs. There are clear patterns within the data, and we have been able to link these patterns to poor practice. While the individual domains being flagged are not an indication of a specific policy or good practice breach, as a collection they are telling.

\textbf{\textit{However}, it is difficult to directly compare different organisations, and with the current data set we are unlikely to be able to prove a difference between the Control and Treatment organisations.} The fact that the Control organisations so consistently visit less malicious and grey domains is troubling to the hypothesis that the digital hygiene efforts improve digital hygiene practices of the CSOs. While this connection seems to make sense, it is not immediately apparent within the data set, and further research is needed.

\subsection{Next Steps}

Now we are familiar with the data, the next step is to look at methods of analysing it to test the hypothesis that the effects of the digital hygiene efforts can be measured by DNS data.

\subsubsection{Time series analysis}

At present it does not seem possible to differentiate the Control and Treatment organisations simply based on the number or proportion of malicious or grey domains visited. One approach we would like to take is to observe organisations as they undergo the digital hygiene treatment: thus switching from a Control organisation to a Treatment organisation. 

At present we only have consistent data between April and November 2018. Subsequent data exists, and crucially we believe this further range includes organisations undergoing this transition. Observing this change would allow us to look for causal inference, to spot whether the intervention had effects on the organisation, and what these effects are. This could be done using Bayesian structural time-series, and Google produce package for this kind of analysis \cite{causalimpact}. We believe that if this data is available, this is the most promising avenue to test our hypothesis.

\subsubsection{Identifying malicious domains}

Another difficulty with the experiment was the messy nature of the threat intelligence feeds. Our initial analysis within this report is based on these feeds, and manual inspection reveals that some of the entries on this list are questionable. 

One solution to this would be if we can implement a systematic solution for detecting malicious domains using DNS, and there is an established body of academic work in this area. In 2011 Antonakakis et al. created a method for automating the detection of new malware domains with a low false positive rate \cite{antonakakis2011detecting}. They have also looked at how to predict fast flux domain and other domains generated by Domain Generation Algorithms \cite{antonakakis2012throw}. Around the same time Stalmans and Irwin made a systematic attempt to examine several different approaches to detect malware domains \cite{stalmans2011framework}. If these approaches could be applied to our dataset to implement a systematic process of identifying malicious domains we could avoid the problem of the unreliability of the treat intelligence feed.\footnote{This problem with the reliability of threat intelligence feeds emphasises a point that may be worth exploring: while cyber threat intelligence is growing in the business world there is little academic discussion regarding a) how to best use it, or b) how effective can it be. Given the time and money that is being spent on threat intelligence feeds and systems that rely on them these questions are worth addressing. While this does not address the primary hypothesis discussed in this paper, this data set does provide an opportunity to explore these follow on questions.}

\section{Conclusion}

This is a preliminary report looking at the DNS data of several CSO to determine whether this information alone is sufficient to prove the efficacy of digital hygiene efforts. In this report we have described the experimental setup and data available to us, and made early attempts to look for signs of poor digital hygiene practices. The results have been cross-referenced against a list of unwanted or malicious domains gathered from open source threat intelligence feeds and some domains have been investigated manually.

Although this report is based on data taken between April and November 2018, this is only an initial evaluation to assess the possibility of using DNS data to evaluate the success of digital hygiene efforts, including education, audit and remediation . This data set is growing over time, and follow on analyses will take place using data from a greater time window.

As well as looking at the data we have discussed the general concept of using DNS to determine the overall state of digital hygiene. Given the signs of malware present in the available data, we believe that such an approach has merit, and is worthy of further analysis.

This report has four key findings as follows. Further breakdown of these issues can be found in section 4.

\begin{itemize}

\item The data set gives an indication of digital hygiene practices, and consistent patterns can be observed over the whole period. 

\item \textit{However}, much of the Threat Intelligence flagging is questionable without manual investigation.

\item Is systematic quantitative evaluation useful without manual investigation? Probably.

\item \textit{However}, it is difficult to directly compare different organisations, and with the current data set we are unlikely to be able to prove a difference between the Control and Treatment organisations.

\end{itemize}

Our work in this introductory report has been limited, and it would be beneficial to explore scaling this approach via automation and statistical modelling. Of particular interest is the possibility of observing domains as they transition from the Control group to the Treatment group. It may also be possible to implement a systematic process for detecting possible malicious domains, removing the need to rely on threat intelligence feeds.

\printbibliography

\end{document}